\documentclass{article}
\usepackage[left=1in, right=1in]{geometry}
\usepackage{authblk}
\usepackage{amsfonts}
\usepackage{mathrsfs}
\usepackage{amsmath}
\usepackage{bm} 
\usepackage{gensymb}
\usepackage{graphicx}
\usepackage{sidecap}
\usepackage{enumerate}
\usepackage{color}
\usepackage[small]{titlesec}
\usepackage{cite}
\begin{document}

\title{Two-photon X-ray Ghost Microscope}
\author[1]{Thomas A. Smith}
\author[2]{Zhehui Wang}
\author[1]{Yanhua Shih}
\affil[1]{Department of Physics, University of Maryland, Baltimore County, Baltimore, MD 21250, USA}
\affil[2]{Los Alamos National Laboratory, Los Alamos, NM 87545, USA}

\date{}
\maketitle

\begin{abstract}
	
	X-ray imaging allows for a non-invasive image of the internal structure of an object. The most common form of X-ray imaging, projectional radiography, is simply a projection or ``shadow'' of the object rather than a point-to-point image possible with a lens. This technique fails to take advantage of the resolving capabilities of short-wavelength X rays. Various X-ray microscopes, typically operating with soft X rays ($< 10$ keV), use focusing X-ray optics to obtain higher resolution images of the internal structure of an object. Due to the short focal length of focusing X-ray optics, you lose the ability to magnify the high-resolution images of the internal structure of thicker objects because the object plane is further away from the focal plane. Here we present an imaging mechanism that utilizes two-photon X-ray ghost imaging to produce a true point-to-point image of the internal structure of an object, with the potential to introduce focusing X-ray optics or a scintillator-lens system to produce a magnified secondary ghost image. The focusing X-ray optics would image the primary ghost image (which has no physical structure to it) allowing the imaging of internal structures deeper than a standard X-ray microscope would be practical for. In principle, once some experimental barriers are overcome, this X-ray ``ghost microscope'' may achieve nanometer spatial resolution and open up new capabilities that would be of interest to the fields of physics, material science, and medical imaging.
\end{abstract}

\section{Introduction}

In classic imaging setups, focusing optics such as imaging lenses play a critical role in producing a diffraction-limited point-to-point correlation between the object plane and the image plane, forming a magnified or demagnified image of the object \cite{BornBook, HechtBook}. If desired, an additional lens system is then able to map the primary image onto a secondary image plane for further magnification or demagnification; notably making an optical microscope possible. Such optical microscopes are commonly used to obtain a high-resolution images of a detailed surface structure of an object. Unlike visible light, X rays can pass through many materials allowing X-ray imaging devices to image the internal structure of an object, where standard projectional radiography is the most common technique \cite{XrayBook, MedicalBook}. If our goal is to obtain higher resolution images of the detailed internal structure of an object or material, the development of an X-ray microscope is necessary but difficult. The first difficulty faced was that traditional lenses are not practical to use because the refractive index for high-energy X rays is always close to 1 \cite{XrayBook}. To overcome this, some alternative focusing X-ray optics have been developed in recent years such as compound refractive lenses, focusing mirrors, and zone plates \cite{SchroerLens, MimuraMirror, ChaoZonePlate}. While zone plates and focusing mirrors typically limited to soft X rays ($< 10$ keV), compound refractive lenses can be designed for hard X rays ($> 10$ keV). Unlike projectional radiography, which is a projection, or ``shadow,'' of the X rays that pass through the object, focusing X-ray optics produce high-resolution images by taking full advantage of the short wavelength of soft X rays; with recent demonstrations producing 10 nm resolution images \cite{ChaoZonePlate}. This is significantly better resolution than projectional radiography technology commonly used for medical purposes with a resolution around 100 $\mu$m \cite{XrayBook, MedicalBook}, and also better than state-of-the-art projectional radiography technology used in research environments (around 700 nm) \cite{NanoReview}. Unfortunately, focusing X-ray optics for both soft and hard X-rays suffer other limitations such as advanced fabrication techniques and experimental constraints such as absorption and low numerical apertures. Due to this, the X-ray imaging technique most commonly used in practice is projectional radiography.

Even as focusing X-ray optics are made more readily available, it is difficult for classic imaging devices such as these to obtain magnified, high-resolution images of the interior structure of a thicker object. To take advantage of the resolving power of short-wavelength X rays by provide significant magnification, the object plane should be near the focal plane. For object planes further away from the focal plane, the magnification may not be enough for a sensor array to capture the desired resolution. In this article, we apply the physics of two-photon ghost imaging to produce a sub-nanometer resolution, lensless, image-forming correlation between the image plane and object plane, for which X rays allow imaging of the internal structure of the object \cite{2005, 2006, Meyers, SmithXRGI}. Through the help of a secondary imaging device, either focusing X-ray optics or a scintillator paired with a visible-light lens assembly, the primary lensless ghost image can be mapped onto a secondary image plane with significant magnification to be resolvable by a standard CCD or CMOS; namely, an X-ray ghost microscope. Similar to lensless ghost imaging in the visible spectrum, the X-ray lensless image-forming correlation is the result of two-photon interference: two randomly created and randomly paired photons interfering with the pair itself \cite{2006, ShihBook}. Lensless ghost imaging is ideal for X-ray imaging because it does not require a lens to produce a primary ghost imaging but also, due to the non-local nature of ``ghost'' imaging, an additional imaging device with limited angular resolution can be placed as close as desired to the primary ghost image plane, which corresponds to a ``sliced'' object plane inside the object. While the use of focusing X-ray optics or the scintillator-lens system may remove the potential for sub-nanometer resolution, this setup allows for high-resolution imaging deep within a thicker object.

\begin{figure}[t]
	\centering
	\includegraphics[width=100mm]{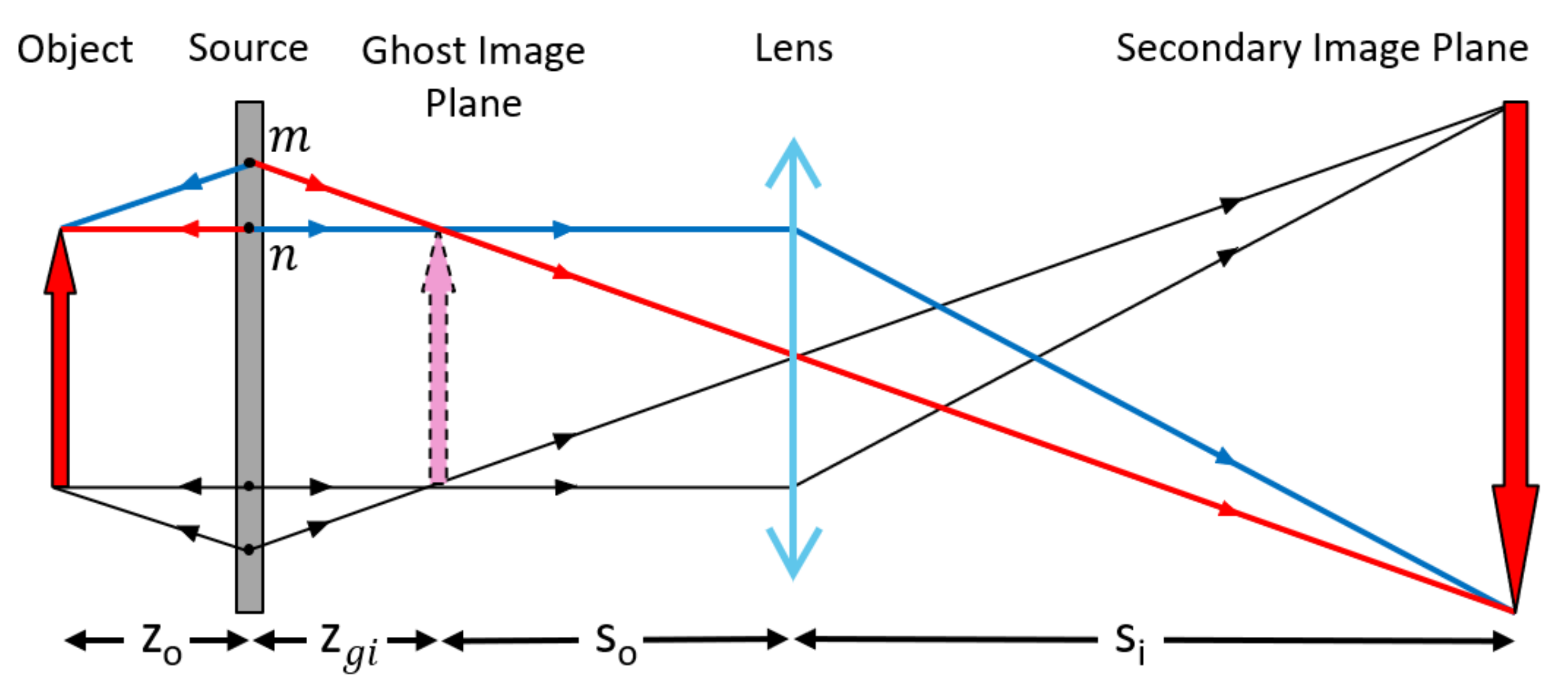}
	\caption{Experimentally achieved with a beam splitter, this ``unfolded'' schematic of the 2005 thermal light ghost imaging demonstration of Valencia \emph{et al.} helps depict the symmetry of the two-photon amplitudes. The one-to-one lensless ghost image and the magnified secondary ghost image are results of two-photon interference. Of the many two-photon amplitudes, such as the red and blue two-photon amplitudes of the random $m$th and $n$th pair of photons shown here, each superpose constructively at a corresponding $\vec{\rho}_{i}$ to from an image.}
	\label{fig:Secondary}
\end{figure}

Visible-light ghost imaging was first demonstrated in 1995 using two-photon interference from entangled photon pairs through the measurement of the second-order coherence function \cite{1995}. 10 years later, it was demonstrated that two-photon interference of randomly created and randomly paired photons in a thermal state can produce a similar point-to-point image-forming correlation without the need of an imaging lens, commonly called lensless ghost imaging \cite{2005, 2006, Meyers, SmithXRGI}. Interestingly, the first lensless ghost imaging of thermal light was observed in the secondary ghost imaging plane with the help of a lens. An unfolded schematic diagram of the 2005 ghost imaging experiment of thermal light is illustrated in Fig.~\ref{fig:Secondary}. X-ray ghost imaging has now been demonstrated multiple times \cite{YuXR,PellicciaXR1,SchoriXR,ZhangXR,PellicciaXR2,CeddiaXR,XrayTomography,KimXR}; however, it is still a developing field. One commonly used technique for current X-ray ghost imaging is to introduce a spatially varied material in the X-ray beam to produce an artificial ``speckle'' pattern. This results in a classical speckle-to-speckle ghost imaging with resolution dependent on the speckle size. However, this ghost imaging technique differs from the mechanism of the diffraction-limited two-photon ghost imaging experiments. Classified according to their experimental setup and working mechanism, we find two classes of ghost imaging: (I) The observed ghost image is produced from a natural point-to-point image forming correlation that is the result of two-photon interference. This class of ghost imaging follows the mechanism of the entangled state ghost imaging of Pittman \emph{et al.} \cite{1995} and the thermal light ghost imaging of Valencia \emph{et al.} \cite{2005}. With thermal light ghost imaging, the point-to-point image-forming correlation is able to achieve spatial resolution of $\lambda / \Delta \theta_s$, where $\Delta \theta_s$ is the angular diameter of the light source and has the potential to be turbulence-free \cite{ShihBook, Meyers2, TFDSIprl, TFDSIoe}. Pelliccia \emph{et al.} appear to use type (I) ghost imaging in their first demonstration of X-ray ghost imaging \cite{PellicciaXR1}. Alternatively, (II) the observed ghost image is produced from a classical speckle-to-speckle correlation shown with visible light \cite{Boyd, Gatti, Gatti2} and with a variety of X-ray sources \cite{SchoriXR,PellicciaXR2,CeddiaXR,XrayTomography,KimXR}. Here, a ghost image of the object is obtained from the coincidences between two sets of identical, artificial ``speckles'' formed by either spatially correlated laser beams or an aperture mask following the light source to produce shadows and bright spots distributed on the object plane and on the ghost image (detector) plane. Included as a classical speckle-to-speckle correlation is computational ghost imaging which removes the need of a beam splitter by measuring the object plane and ghost image plane at separate times with an identical speckle pattern \cite{YuXR, ZhangXR, CompGI}. These artificial speckles are different from the two-photon interference that produces a point-to-point image-forming correlation and, correspondingly, this type of ghost imaging has a resolution dependent on the size of the classically formed speckles. Through the correlation measurement, this type of ghost imaging observes a projection or a shadow of the object, comparable to how classic X-ray imaging technology is a projection of the object. Both classes of ghost imaging use the measurement of intensity fluctuation correlation to obtain an image, utilizing either changes in the artificial speckle distribution over time or fluctuations resulting from two-photon interference.

The X-ray ghost microscope studied in this article belongs to type (I) ghost imaging via two-photon interference. As with lensless ghost imaging, the spatial resolution of X-ray ghost imaging before magnification is determined by $\lambda / \Delta \theta_s$, where $\lambda$ is the wavelength and $\Delta \theta_s$ is the angular diameter of the X-ray source. This unique characteristic of lensless ghost imaging makes the resolving potential of an X-ray microscope a point of interest. As an example, using a high-energy ($>$20 keV) X-ray source with a relatively large angular diameter may produce a one-to-one (no magnification or demagnification) lensless ghost image with sub-nanometer resolution. It should be noted that this is the opposite of what is desired for classic projectional X-ray imaging which sees an increase in resolution when using a smaller, point-like source, such as modern X-ray tubes and synchrotron X-ray sources \cite{XrayBook, footnoteXFEL}. Type (I) ghost imaging is still possible with these small sources, but the resolution will be reduced. When expecting a ghost image with even sub-micrometer resolution, the image becomes unresolvable by any state-of-the-art CCD or CMOS detector arrays. However, this primary ghost image can be magnified significantly onto a secondary ghost image plane through either focusing X-ray optics or a scintillator-lens system, potentially making nanometer sized features resolvable by a CCD or CMOS with micrometer pixels. Importantly, the secondary ghost image is produced directly by the point-to-point image-forming correlation between the object plane, $\vec{\rho}_o$, and the secondary ghost image plane, $\vec{\rho}_i$, where $\vec{\rho}_o$ and $\vec{\rho}_i$ represent the transverse coordinates of the object plane and image plane, respectively. This peculiar feature preserves the diffraction-limited spatial resolution from two-photon interference. An additional benefit of two-photon interference is that this X-ray ghost microscope can be set up such that it is insensitive to any rapid phase variations along the optical path due to random changes in composition, density, length, index of refraction, or medium vibration, namely ``turbulence-free'' \cite{Meyers2, TFDSIprl, TFDSIoe}. Overcoming vibrations is especially important for the extremely high resolution imaging obtainable with the X-ray ghost microscope. 

This article is organized in four sections: (1) we introduce the concept of two-photon interference induced intensity fluctuation correlation, (2) we show a point-to-point image-forming correlation for producing a lensless ghost image, (3) we extend this ghost image to a secondary image using either focusing X-ray optics or a scintillator-lens system, and (4) we discuss how to obtain observable intensity fluctuation correlation when using an X-ray source with a broad spectrum. 

\section{Two-photon interference induced X-ray intensity fluctuation correlation}

Recall, in 1905, Einstein introduced a granularity to radiation, abandoning the continuum interpretation of Maxwell \cite{Einstein-1905}. This led to a microscopic picture of radiation and a statistical view of light, including X rays. In Einstein's picture, a light source consists of many point-like sub-sources, each of which emit their own subfields. Originally labeled in German by ``strahlenb\"undel,'' which translates to ``bundle of ray'' in English, we now label these bundles of light as subfields or the quantum term of photons. For a thermal light source, these subfields (photons) are emitted in a random manner such that the $m$th subfield emitted from the $m$th point-like sub-source may propagate in all possible directions with a random initial phase. It has been shown that an effective wavefunction can be defined from the quantum theory of optical coherence \cite{Glauber1, Glauber2} to specify the space-time behavior of a photon \cite{ShihBook, ScullyBook}. The effective wavefunction of a photon in a thermal state is mathematically the same function as Einstein's subfield. In Einstein's picture, the radiation measured at coordinate $(\bm{r}, t)$ is the result of a superposition of a large number of subfields,
\begin{align}
E(\bm{r}, t) = \sum_{m}E_m(\bm{r}, t) 
= \sum_{m}E_m(\bm{r}_m, t_m) \, g_m(\bm{r}_m, t_m; \bm{r}, t),
\end{align}
where $E_m(\bm{r}_m, t_m)$ labels the subfield emitted from the $m$th sub-source at coordinate ($\bm{r}_m, t_m$), and $g_m(\bm{r}_m, t_m; \bm{r}, t)$ represents the field propagator or Green's function that propagates the $m$th subfield from coordinate $(\bm{r}_m, t_m)$ to coordinate $(\bm{r}, t)$. Shortening the notation, we will replace $E_m(\bm{r}_m, t_m)$ with $E_m$ and $g_m(\bm{r}_m, t_m; \bm{r}, t)$ with $g_m(\bm{r}, t)$. In Einstein's picture of light, the expectation value of the intensity corresponds to a statistical ensemble average which takes into account \emph{all possible realizations of the field} or, more specifically, takes into account \emph{all possible relative phases of the subfields}:
\begin{align}\label{Image-00}
\langle I(\bm{r}, t) \rangle & = \langle E^*(\bm{r}, t) E(\bm{r}, t) \rangle \cr
& = \big{\langle} \sum_{m} E^*_m(\bm{r}, t) \sum_n E_n(\bm{r}, t) \big{\rangle} \cr
& = \big{\langle} \sum_{m} \big{|} E_m(\bm{r}, t) \big{|}^2 \big{\rangle}
+ \big{\langle} \sum_{m\neq n} E^*_m(\bm{r}, t) E_n(\bm{r}, t) \big{\rangle} \cr
& = \sum_{m} \big{|} E_m(\bm{r}, t) \big{|}^2.
\end{align}
The expectation value of $\sum_{m\neq n} E^*_m(\bm{r}, t) E_n(\bm{r}, t)$ goes to zero when taking into account all possible relative phases of the subfields. We may conclude the theoretical expectation value of intensity $\langle I(\bm{r}, t) \rangle = \sum_{m} \big{|} E_m(\bm{r}, t) \big{|}^2$ is the result of the $m$th subfield interfering with the $m$th subfield itself, while the $m\neq n$ term is a result of the $m$th subfield interfering with a different $n$th subfield. In a realistic measurement in which only a limited number of subfields contribute to the measurement, all possible phases may not be present, meaning the $m\neq n$ term may not vanish and contribute noise to the measurement. We may name this term a two-photon interference induced intensity fluctuation,   
\begin{align}\label{IntensityFluctuation-00}
\Delta I(\bm{r}, t) = \sum_{m\neq n} E^*_m(\bm{r}, t) E_n(\bm{r}, t). 
\end{align}
The two-photon induced intensity fluctuation is different from other traditional ``intensity fluctuations.'' For instance, the number of subfields contributing to a measurement may vary from measurement to measurement and thus the value of $\sum_{m} \big{|} E_m(\bm{r}, t) \big{|}^2$ may change from measurement to measurement. These variations may be from a classical mask in the path of the light or variations in the intensity of the source. These would be the fluctuations used in type (II) ghost imaging. Note that in the following discussion, no classical intensity fluctuations are involved, neither from a spatial intensity distribution nor temporal intensity distribution, but only taking the two-photon induced intensity fluctuations of thermal light.  

The X-ray ghost microscope studied in this article utilizes the two-photon induced intensity fluctuation correlation $\langle \Delta I(\bm{r}_1, t_1) \Delta I(\bm{r}_2, t_2) \rangle$ for type (I) ghost imaging. Although the expectation or ensemble average of the intensity fluctuation measured by a single detector $D_j$, $j = 1, 2$, is zero, $\langle \Delta I(\bm{r}_j, t_j) \rangle = \langle \sum_{m\neq n} E^*_m(\bm{r}_j, t_j) E_n(\bm{r}_j, t_j) \rangle = 0$, the expectation or ensemble average of the correlation of the intensity fluctuations measured by $D_1$ and $D_2$, jointly, may not equal zero, 
\begin{align}\label{fluctuation-22}
\langle \Delta I(\bm{r}_1, t_1) \Delta I(\bm{r}_2, t_2) \rangle & = {\Big{\langle}} \sum_{m\neq n} E_m^*(\bm{r}_1, t_1) E_{n}(\bm{r}_1, t_1)
\sum_{p\neq q} E_p^*(\bm{r}_2, t_2) E_{q}(\bm{r}_2, t_2) {\Big{\rangle}} \cr
& = \sum_{m\neq n} E_m^*(\bm{r}_1, t_1) E_{n}(\bm{r}_1, t_1)
E_n^*(\bm{r}_2, t_2) E_{m}(\bm{r}_2, t_2).
\end{align}
Due to the random relative phases between the subfields canceling in a specific case, when $m = q$ and $n = p$, there is a surviving term in the above summation. Mathematically, the result of Eq.~(\ref{fluctuation-22}) can be represented as the cross term of the following superposition,
\begin{align}\label{two-photon-interference-00}
G^{(2)}(\bm{r}_1, t_1; \bm{r}_2, t_2) & = \sum_{m \neq n} \big{|} E_m(\bm{r}_1, t_1) E_n(\bm{r}_2, t_2)
+ E_n(\bm{r}_1, t_1) E_m(\bm{r}_2, t_2) \big{|}^2 \cr
& = \langle I(\bm{r}_1, t_1) \rangle \langle I(\bm{r}_2, t_2) \rangle + \langle \Delta I(\bm{r}_1, t_1) \Delta I(\bm{r}_2, t_2) \rangle
\end{align}
corresponding to the superposition of two different yet indistinguishable alternatives of joint photodetection: (1) the $m$th subfield (photon) is measured at $(\bm{r}_1, t_1)$ while the $n$th subfield (photon) is measured at $(\bm{r}_2, t_2)$; (2) the $n$th subfield (photon) is measured at $(\bm{r}_1, t_1)$ while the $m$th subfield (photon) is measured at $(\bm{r}_2, t_2)$. The cross term of this superposition models the concept of randomly paired photons interfering with the pair itself, namely two-photon interference. 

\section{Lensless ghost imaging of X-ray} \label{XRGI}

\begin{figure}
	\centering
	\includegraphics[width=60mm]{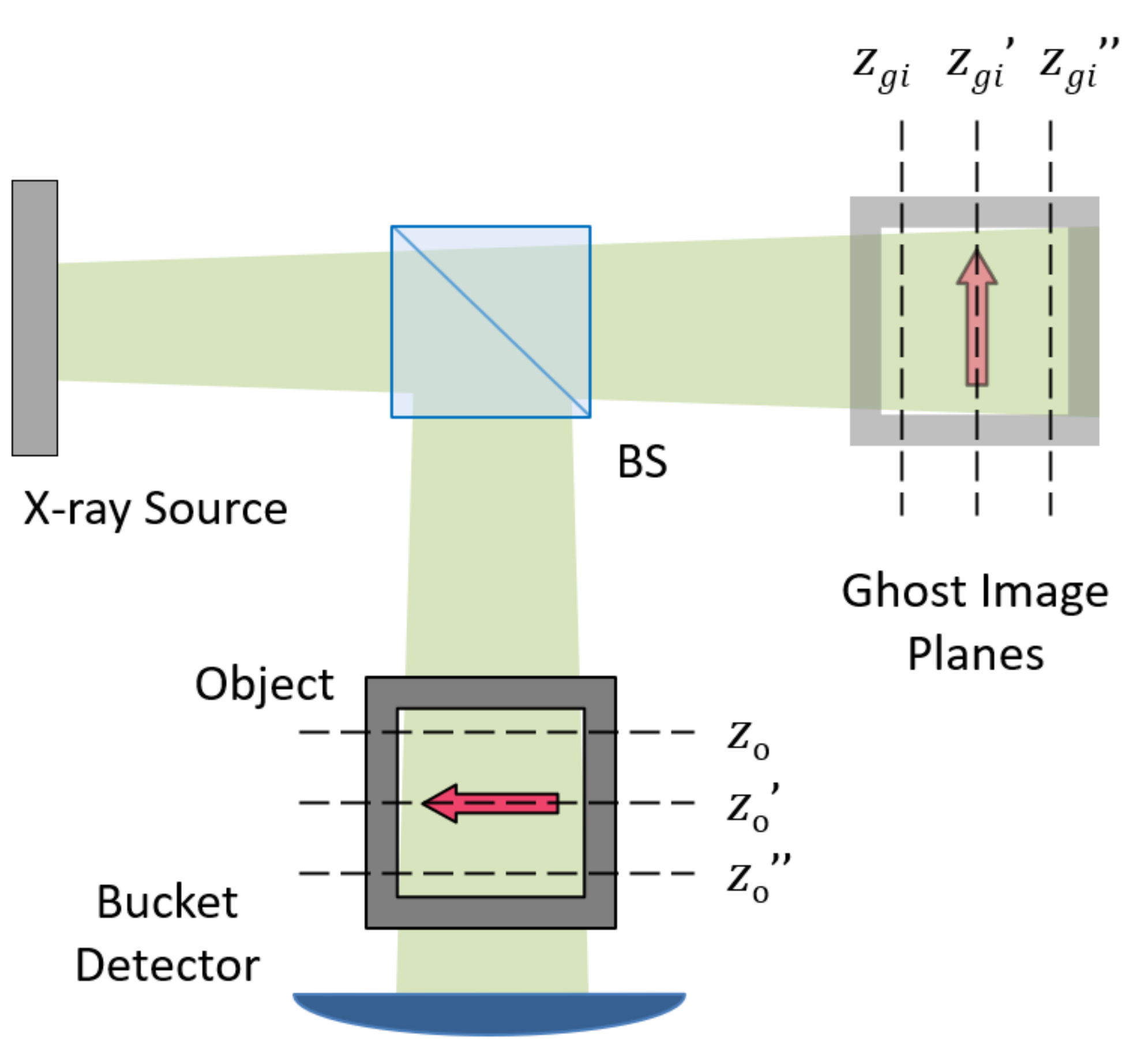}
	\caption{X-ray ghost imaging. A beam splitter (most likely a crystal aligned to utilize Laue diffraction which would not provide 90\degree\ separation as depicted) creates two paths for the beam, one directed passing through the object and followed by a bucket detector, and the second directed at a 2-D photodetection array (CCD or CMOS). The array is placed on the desired ghost image plane, $z_1 = z_{gi}$, corresponding to different ``slices'', or cross sections inside the object: $z_{gi} = z_o, z_{gi}' = z_o', z_{gi}'' = z_o'',$ etc.}
	\label{fig:XRGI}
\end{figure}

To better visualize the X-ray ghost microscope it is best to start with the working mechanism of X-ray lensless ghost imaging. This simple experimental setup, schematically illustrated in Fig.~\ref{fig:XRGI}, consists of an X-ray beamsplitter \cite{footnoteBS} that divides the X-ray beam from a disk-like source into two beams. A 2-D array of X-ray detectors, $D_1$, is placed in beam-one at a selected plane of $z_1 = z_{gi}$ in the Fresnel near-field. Following the output of beam-two, we place an object followed by a bucket X-ray photodetector, $D_2$, which collects all X rays transmitted from the object.
To calculate the X-ray ghost image forming correlation, we apply the Fresnel near field propagator, or Green's function, to propagate the field from one space-time location, $(\vec{\rho}_n, z_n)$, to another space-time location at $(\vec{\rho}_j, z_j)$ \cite{ShihBook}. 
\begin{align}\label{Green-11}
g_n(\vec{\rho}_j, z_j) = \frac{c_o}{|z_j - z_n|} e^{-i \omega \tau_j} e^{i \frac{\omega}{2 c |z_j - z_n|} |\vec{\rho}_j -\vec{\rho}_n |^2 },
\end{align}
where $c_0$ is a normalization constant and $\tau_j \equiv t_j - |z_j - z_n|/c$. Due to the use of short wavelength X rays, higher-order approximation, or a numerical computation, may be necessary instead of a simple Fresnel propagator. However, it is a mathematical conclusion that higher-order approximations will produce a ``narrower'' correlation than that of the Fresnel. To simplify the mathematics, we will keep the Fresnel approximation for this proof-of-principle analysis. Also, for the following discussion, we will assume a monochromatic source and will address concerns of temporal coherence in Section~\ref{temporal}. Assuming a disk-like source and randomly distributed and randomly radiated point-like sub-sources, we can approximate the sum of $m$ in Eq.~(\ref{fluctuation-22}) into an integral of $\vec{\rho}_s$ on the source plane of $z_s = 0$. Before considering the detectors, we can look at the object plane on beam-two, $z_o$, and the corresponding plane on beam-one, which we will label $z_{gi}$. The result of intensity fluctuation correlation is
\begin{align}\label{HBT-near-field-22}
\langle \Delta I(\vec{\rho}_{gi}, z_{gi}) \Delta I(\vec{\rho}_{o}, z_o) \rangle \big{|}_{z_{gi} = z_o} & \propto \Big{\langle} \Big{|} \int d \vec{\rho}_s \,\big{[} g^*_s(\vec{\rho}_{gi}, z_{gi}) \big{]} \big{[} g_s(\vec{\rho}_o, z_o) \big{]} \Big{|}^2 \Big{\rangle} \cr
& \propto \textrm{somb}^2 \frac{\pi \Delta \theta_s}{\lambda} |\vec{\rho}_{gi} - \vec{\rho}_{o}| 
\end{align}
where the somb-function is defined as $2J_1(x)/x$, where $J_1(x)$ is the Bessel function  of the first kind, and $\Delta\theta_s \approx 2R/z_{gi}$ is the angular diameter of the radiation source. This point-to-spot correlation is similar to that of the original Hanbury Brown-Twiss experiment that started the practice of optical correlation measurements, except it is measured in the Fresnel near-field instead of the Fraunhofer far-field \cite{HBT1, HBT2, HBT3}. This point-to-spot correlation has also been demonstrated at X-ray synchrotron sources \cite{XrayHBT1997, XrayHBT1999, XrayHBT2001}. Due to the high energy (short wavelength) nature of X-rays, this point-to-spot correlation is much more narrower than that of the visible-light. For instance, a $\sim 0.5 \times10^{-10}$ meter wavelength ($\sim 25$ keV) X-ray source with angular diameter of $>10^{-1}$ rad may achieve sub-nanometer correlation. We can approximate this point-to-spot correlation as a point-to-point correlation between $\vec{\rho}_{gi}$ and $\vec{\rho}_o$. This point-to-point correlation, or ghost image forming function, identifies a unique internal plane of the object, $z_{gi} = z_o$, with the X-ray ghost image observed from the X-ray intensity fluctuation, or photon number fluctuation, correlation. With the point-to-point approximation and $z_{gi} = z_o$, we may approximate the X-ray intensity fluctuation correlation as
\begin{align}
\langle \Delta I(\vec{\rho}_{gi}, z_{gi}) \Delta I(\vec{\rho}_{o}, z_o) \rangle
\propto \delta (z_{gi} - z_o) \, \delta (|\vec{\rho}_{gi} - \vec{\rho}_{o}|).
\end{align}
We now calculate the X-ray lensless ghost image by including a 2-D photodetection array on the $z_1 = z_{gi}$ plane and a bucket detector, $D_2$, which integrates all possible X rays transmitted through each transverse coordinate point $\vec{\rho}_o$ of the object. The internal ``aperture function'' of an object cannot be simply represented as a 2-D function $A(\vec{\rho}_{o}, z_o)$ as typically done for visible-light imaging. In visible light imaging, $A(\vec{\rho}_{o})$ is usually used to represent the surface plane of an object; however, for X-ray ghost imaging it is reasonable to model a 3-D aperture function representing the internal structure of an object,
\begin{align}
A(\vec{\rho}_{o}) \approx \int d z_o A(\vec{\rho}_{o}, z_o).
\end{align}
Assuming perfect temporal correlation is satisfied experimentally, the intensity fluctuation correlation results in
\begin{align}
\langle \Delta I(\vec{\rho}_{1}, z_1) \Delta I_2 \rangle
& \propto \Big{|} \int d \vec{\rho}_{o} \int d z_o \, A(\vec{\rho}_{o}, z_o) \delta (z_1 - z_o)\, \textrm{somb} \frac{\pi \Delta \theta_s}{\lambda} |\vec{\rho}_{1} - \vec{\rho}_{o}| \Big{|}^2 \cr
& \approx \Big{|} \int d \vec{\rho}_{o} \int d z_o \, A(\vec{\rho}_{o}, z_o) \delta (z_1 - z_o) \, \delta (|\vec{\rho}_{1} - \vec{\rho}_{o}|) \Big{|}^2 \cr
& \approx \big{|} A(\vec{\rho}_{1} = \vec{\rho}_{o}, z_1 = z_o) \big{|}^2,
\end{align}
indicating the reproduction of a 2-D X-ray ghost image of the internal transverse cross section of $z_o = z_{gi} = z_1$. In other words, the longitudinal position of the 2-D X-ray detector array, $z_1$, selected an object plane, $z_o = z_1$, of the internal structure of the object to image. Interestingly, when the 2-D X-ray detector array is linearly scanned from $z_1$ to $z'_1$ to $z''_1$ along the optical axis, the selected object plane will be changed from $z_o = z_1$ to $z'_o = z'_1$ to $z''_o = z''_1$, respectively, as illustrated in Fig.~\ref{fig:XRGI}. The imaging resolution along the $z$-axis is the coherence time of the light, similar to optical coherence tomography \cite{OCT}. By scanning the position of the 2-D X-ray detector array, a set of slices of the internal structure of the object can be grouped together to form a 3-D ghost image of the object with sub-nanometer resolution. This differs from traditional X-ray computerized tomography (CT) imaging \cite{CT} and ghost tomography (GT) demonstrated by Kingston \emph{et al.} \cite{XrayTomography} which rely on rotating the object or revolving the detectors around the object 360\degree \cite{footnote3D}.

\section{X-ray Ghost Microscope}

Unfortunately, the sub-nanometer resolution of lensless ghost imaging is unresolvable by any state-of-the-art 2-D photodetector array, such as CCD or CMOS sensors which may have micrometer sized pixels, unless photodetector array technology ever advances to sub-nanometer-sized pixels. In order to resolve the two-photon X-ray ghost image with a standard 2-D photodetector array, significant magnification is necessary. 

\subsection{Configuration I}

\begin{figure*}[ht]
	\centering
	\includegraphics[width=125mm]{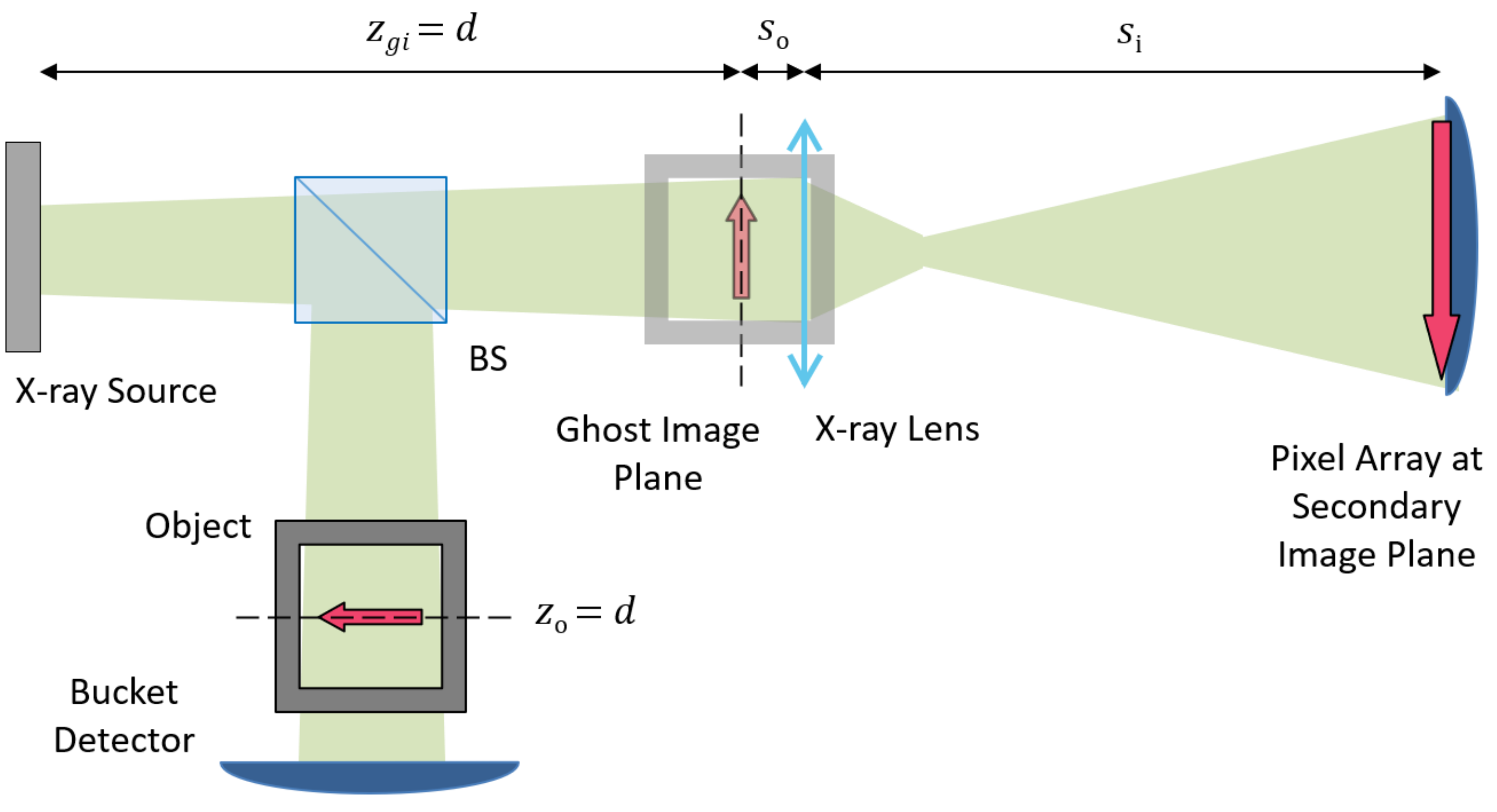}
	\caption{The X-ray ghost microscope. Focusing X-ray optics are used to produce a magnified secondary ghost image prior to the 2-D photodetection array which allows for a high resolution image of the object to be resolvable by a standard CCD or CMOS. Different primary ghost image planes correspond to different ``slices'' of the object. By scanning the lens and $D_1$ along the optical axis, we select new $z_{gi} = z_o = d$ planes to be in focus, obtaining a magnified 3-D image of the internal structure of the object.}
	\label{fig:XRGM1}
\end{figure*}

To magnify the X-ray ghost image to a secondary ghost image plane we may follow the optical setup demonstrated by Valencia \emph{et al.} in 2005 for visible light ghost imaging. The schematic design of this standard X-ray ghost microscope matches what an optical setup would be and is shown in Fig.~\ref{fig:XRGM1}. For this configuration of the X-ray ghost microscope, focusing X-ray optics such as compound refractive lenses, X-ray zone plates, etc. can be used to produce a magnified secondary ghost image \cite{SchroerLens, MimuraMirror, ChaoZonePlate}. Note that focusing X-ray optics may limit the usable energy levels of the X-ray source unless the technology advances to accommodate higher energy levels. It should also be stated that the use of focusing X-ray optics to form a secondary ghost image is only possible with the true image formation of type (I) ghost imaging. If you ``unfold'' the paths of light in type (II) ghost imaging, as done with type (I) ghost imaging in Fig.~\ref{fig:Secondary}, you see that the image formation more closely resembles a projection or ``shadow'' which is not something that can be imaged by a lens. 
As with the primary ghost image, this secondary ghost image is observable from the measurement of the intensity fluctuation correlation. To confirm the nonlocal point-to-point correlation between the object plane, $z_o$, and the secondary image plane of the lens, $z_i$ (where $D_1$ will be placed, $z_1 = z_i$), we adjust Eq.~(\ref{HBT-near-field-22}) to include the lens system,
\begin{align}
\langle \Delta I(\vec{\rho}_{i}, z_i) \Delta I(\vec{\rho}_{o}, z_o) \rangle & \propto \Big{|} \int d \vec{\rho}_s \big{[} g^*_s(\vec{\rho}_o, z_o) \big{]} \big{[} g_s(\vec{\rho}_i, z_i) \big{]} \Big{|}^2 \cr
& \propto \Big{|} \int d \vec{\rho}_s \big{[} g^*_s(\vec{\rho}_o, z_o) \big{]} \big{[} g_s(\vec{\rho}_{gi}, z_{gi}) \int d \vec{\rho}_{gi} \, g_{gi}(\vec{\rho}_L, z_L) \cr
& \ \ \ \ \ \times g_{Lens} \int d \vec{\rho}_L \, g_L(\vec{\rho}_i, z_i) \big{]} \Big{|}^2.
\end{align}
Here $g_s(\vec{\rho}_{gi}, z_{gi})$, $g_{gi}(\vec{\rho}_L, z_L)$, $g_{Lens}$, and $g_L(\vec{\rho}_i, z_i)$ are the Green's functions propagating the field from the source plane to the one-to-one primary ghost image plane, from the primary ghost image plane to the lens plane, from the input plane of the lens to the output plane of the lens, and from the lens plane to the secondary ghost image plane, respectively. This results in,
\begin{align}\label{GhostImage}
\langle \Delta I(\vec{\rho}_{i}, z_i) \Delta I(\vec{\rho}_{o}, z_o) \rangle & \propto \Big{\langle} \Big{|} \int d \vec{\rho}_{gi} \, \delta(|\vec{\rho}_{gi} - \vec{\rho}_o|) \,
\textrm{somb} \frac{\pi D }{s_o \lambda} |\vec{\rho}_{gi} - \vec{\rho}_{i} / \mu | \Big{|}^2 \Big{\rangle} \cr
& \propto \textrm{somb}^2 \frac{\pi D }{s_o \lambda} |\vec{\rho}_{o} - \vec{\rho}_{i} / \mu |,
\end{align} 
indicating a point-to-spot secondary ghost image forming function \cite{2005} \cite{ShihBook}. In Eq.~(\ref{GhostImage}), $D$ is the diameter of the lens, $s_o$ is the distance from the primary ghost image to the lens, $s_i$ is the distance from the lens to the secondary ghost image, satisfying the Gaussian thin lens equation $1/s_o + 1/s_i = 1/f$, where $f$ is the focal length of the lens and $\mu = s_i / s_o$ is the magnification factor of the secondary ghost image. The 2-D detector array, $D_1$, is now placed on the image plane of the lens (secondary ghost image plane), $z_1 = z_i$. The position of $D_1$ defines the position of $z_{gi}$ and thus defines the ``slice'' at $z_o = z_{gi}$ of the internal cross section of the object by means of the Gaussian thin-lens equation. We then take into account the complex aperture function $\tilde{A}(\vec{\rho}_o, z_{o} = z_{gi})$ and the bucket detector $D_2$ for the calculation. A magnified secondary ghost image of the aperture function is then observed from the joint detection between the CCD (CMOS), $D_1$, and the bucket detector $D_2$,
\begin{align}
\langle \Delta I(\vec{\rho}_{1}, z_1) \Delta I_2 \rangle & \propto \Big{\langle} \Big{|} \int d \vec{\rho}_{o} \, \tilde{A}(\vec{\rho}_o, z_o = z_{gi}) \, \textrm{somb} \frac{\pi D }{s_o \lambda} |\vec{\rho}_{o} - \vec{\rho}_{1} / \mu| \Big{|}^2 \Big{\rangle} \cr
& \propto \int d \vec{\rho}_{o} \big{|} \tilde{A}(\vec{\rho}_o, z_o = z_{gi}) \big{|}^2 \textrm{somb}^2 \frac{\pi D }{s_o \lambda} |\vec{\rho}_{o} - \vec{\rho}_{1} / \mu| \cr
& \approx \big{|} \tilde{A}(\vec{\rho}_1 / \mu, z_o = z_{gi}) \big{|}^2.
\end{align}
Scanning $D_1$ from one position to another along the optical axis, or refocusing the microscope from one ghost image plane of $z_{gi}$ to $z'_{gi}$ etc., we obtain a magnified 3-D ghost image of the internal structure of the object.

Due to the high resolution of the primary ghost image, the result of the magnified secondary ghost image is identical to if an image from a classical X-ray microscope was obtained with the focusing X-ray optics. With an angular resolution limited by Rayleigh's criterion, the best spatial resolution one can obtain is by putting the object plane near the focal plane (with the potential for nanometer resolution \cite{ChaoZonePlate}). In addition to equivalent resolution, one major benefit to the ghost microscope setup is that the X-ray optics are imaging the ghost image plane, which isn't physically present, as opposed to imaging the physical object (hence the name ``ghost'' microscope). Imaging some objects with a classical X-ray microscope may not be an issue, but often it may be desirable to image the deeper internal structure (e.g. a bone) of a thicker object (e.g. a body) for which the focusing X-ray optics can not physically be close enough to the internal structure to image it with proper magnification and resolution. This would not be an issue for the X-ray ghost microscope as the physical object is placed ``nonlocally'' on the other path following the beam splitter and is simply followed by a bucket detector. 

\subsection{Configuration II}

\begin{figure*}[hbt]
	\centering
	\includegraphics[width=125mm]{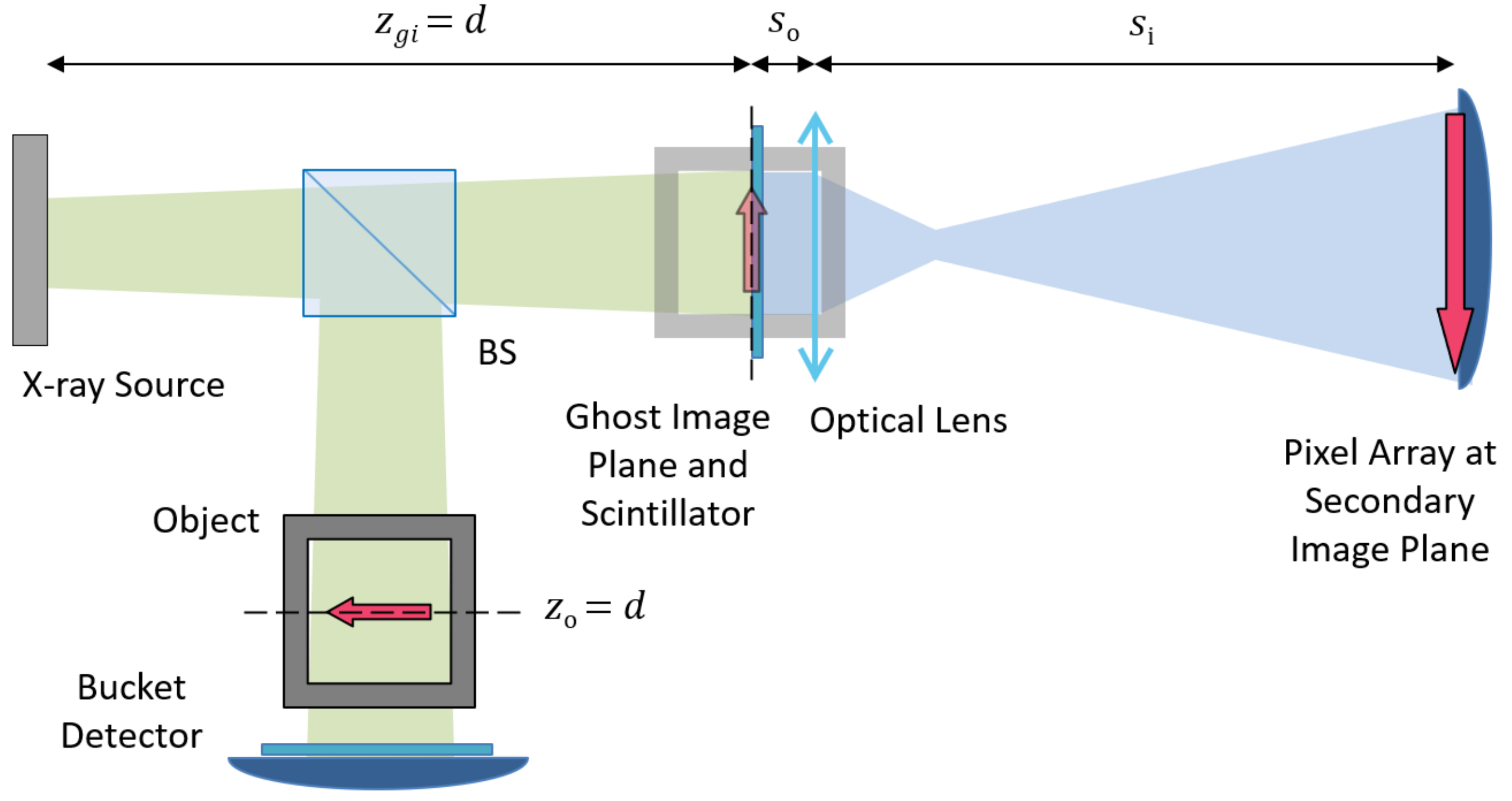}
	\caption{The adjusted X-ray ghost microscope. Nearly identical to the standard X-ray ghost microscope, but now a scintillator is placed on the ghost image plane to convert the X-ray ghost image into the visible spectrum. Now a lens (or lens system) operating in the visible spectrum produces a magnified secondary ghost image.}
	\label{fig:XRGM2}
\end{figure*}

If the use of X-ray optics is not possible or not preferred, one can place a scintillator on the primary ghost image plane, $z_{gi}$ to convert it into the visible spectrum, allowing it to then be magnified by a visible-light lens system onto a secondary image plane. Unlike configuration I, this design is more suitable for higher energy X-ray imaging, such as $\geq 20$ keV, depending on the scintillator used. The schematic design of the adjusted X-ray microscope is shown in Fig.~\ref{fig:XRGM2}.
Unlike the standard X-ray ghost microscope which has a clear path for the two-photon amplitudes from the light source to the detectors, here it is not clear that the scintillator-lens system preserves the result of the two-photon interference, and thus the secondary ghost image. To understand how the secondary ghost image is preserved, we can say that the scintillator essentially ``detects'' the X rays, thus establishing the presence of the X-ray ghost image on the scintillator plane; although, prior to correlation, this plane is simply a distribution of quantum speckles from interfering photon pairs. The scintillator converts these X-ray fluctuations to the visible spectrum, $\Delta I_{\textrm{X}}(\vec{\rho}_{gi}, z_{gi}) \approx \Delta I_{\textrm{v}}(\vec{\rho}_{gi}, z_{gi})$. The lens system then images this distribution of fluctuations and produces a diffraction-limited magnified image of them on the image plane,
\begin{align}
\Delta I(\vec{\rho}_i, z_i) & = \sum_{m \neq n} E^*_m(\vec{\rho}_i, z_i) E_n(\vec{\rho}_i, z_i) \cr
& = \sum_{m \neq n} \big{[} \int_{lens} d \vec{\rho}_l \, E^*_m(\vec{\rho}_{gi}) g^*_{gi}(\vec{\rho}_l, z_l) g_{Lens} g^*_l(\vec{\rho}_i, z_i) \big{]} \cr
& \ \ \ \ \ \times \big{[} \int_{lens} d \vec{\rho'}_l \, E_n(\vec{\rho}_{gi}) g_{gi}(\vec{\rho'}_l, z'_l) g_{Lens} g_{l'}(\vec{\rho}_i, z_i)  \big{]} \cr
& = \sum_{m \neq n} E^*_m(\vec{\rho}_{gi}) E_n(\vec{\rho}_{gi}) \,
\textrm{somb}^2\frac{\pi D}{s_o \lambda_{\textrm{v}}} |\vec{\rho}_{o} + \vec{\rho}_{i}/\mu| \cr
& = \Delta I(\vec{\rho}_{gi}, z_{gi}) \, \textrm{somb}^2\frac{\pi D }{s_o \lambda_{\textrm{v}}} |\vec{\rho}_{o} + \vec{\rho}_{i}/\mu|.
\end{align}
where $\lambda_{\textrm{v}}$ is the center wavelength emitted from the scintillator. It is interesting that the X-ray intensity fluctuations at $(\vec{\rho}_{gi}, z_{gi})$ of the primary ghost image plane are ``propagated'' to an unique point in the secondary image plane $(\vec{\rho}_i, z_i)$, where $(\vec{\rho}_i, z_i)$ is defined by the somb-function and the Gaussian thin-lens equation of the visible-light microscope. Unlike configuration I, this process allows the use of type (II) ghost imaging as well, which would involve ``detection'' and imaging of the classical speckle-pattern. Although type (II) ghost imaging may introduce lower resolution and doesn't not have the capability of linear ghost tomography. Returning to the application of type (I) ghost imaging, a magnified secondary ghost image is observable from the correlation measurement between the intensity fluctuations of the X-ray and the intensity fluctuations of the visible light,
\begin{align}
\langle \Delta I(\vec{\rho}_{1}, z_1) \Delta I_2 \rangle & \propto \Big{\langle} \Big{|} \int d \vec{\rho}_{o} \, \tilde{A}(\vec{\rho}_o, z_o = z_{gi}) \, \textrm{somb} \frac{\pi D }{s_o \lambda_{\textrm{v}}} |\vec{\rho}_{o} - \vec{\rho}_{1} / \mu| \Big{|}^2 \Big{\rangle} \cr
& \propto \int d \vec{\rho}_{o} \, |\tilde{A}(\vec{\rho}_o, z_o = z_{gi})|^2 \, \textrm{somb}^2 \frac{\pi D }{s_o \lambda_{\textrm{v}}} |\vec{\rho}_{o} - \vec{\rho}_{1} / \mu|.
\end{align}
Interestingly, because each ghost image plane along this beam represents a different ``slice'' of the object (as discussed in Section~\ref{XRGI}), a thick scintillator may contain multiple ghost image planes depending on the penetrating depth of the X rays. The resolution along the $z$-axis of the image would be more dependent on the depth of field of the lens system. Regarding the spatial resolution, even with a new dependence on visible light, the most ideal setup of pairing the scintillator with super-resolving visible-light imaging techniques would allow for nanometer resolution \cite{SuperRes}. Of course, more often a standard compound microscope will be the best accessible option with a resolution limit of around 200 nm. However, even with a 200 nm limit, this setup would still have benefits of standard optical microscopy as it allows for imaging the internal structure of the object. It also allows the scintillator to be placed directly on the desired ghost image plane as opposed to behind the physical object in projectional radiography. If preferred, it would also be possible to use X-ray optics prior to a scintillator and an optical lens system following the scintillator. 

One factor that may aid in the high resolving capabilities of the X-ray ghost microscope is the optical stability of two-photon interference. It has been shown that ghost imaging and other two-photon interference phenomena can be set up to achieve turbulence-free measurements \cite{TFDSIprl, TFDSIoe, Meyers2, HBT2}. This is achieved when the superposed two-photon amplitudes experience the same turbulence and medium vibrations along their optical paths, meaning any composition, density, length, refractive index, or medium vibration induced random phase variations along the optical paths do not have any effect on each individual two-photon interference. While optical turbulence may not be prominent in the X-ray regime, this technique also includes small vibrations of the object or detectors that would become more likely to blur a classical image at the resolutions discussed.

\section{Achieving observable two-photon interference with X-ray sources} \label{temporal}

The intensity fluctuation correlation utilized for the X-ray ghost microscope is directly related to the second-order correlation function, $G^{(2)}(\bm{r}_1, t_1; \bm{r}_2, t_2)$, of X-rays in a thermal state \cite{ShihBook,Glauber1,Glauber2,ScullyBook}. This second-order correlation function follows a quantum description of light, so we use the quantized notation of photon numbers, $n(\bm{r}_j, t_j) \propto I(\bm{r}_j, t_j)$ and $\Delta n(\bm{r}_j, t_j) \propto \Delta I(\bm{r}_j, t_j)$ for $j = 1, 2$. Up to this point, we have assumed perfect temporal correlation to allow us to emphasize the spatial portion of this function. Now we will do the opposite and focus on the temporal aspect of the measurement. Similar to ghost imaging with sunlight, we have to face the problem caused by the extremely broad spectrum of X-ray sources and relatively slow detectors. Due to the limited ability of the photodetectors in determining the registration time of a photoelectron, the response time of the photodetectors and the associated electronics will affect the measurement of $G^{(2)}(t_{1}-t_{2})$, where $t_j$ is the registration time of photodetectors $D_1$ and $D_2$. For instance, due to the slow response time of the photodetectors, relatively speaking, the measured second-order correlation may have a much wider temporal width and much smaller amplitude. Note that this also holds true with true with the relatively slow decay time of the scintillator in configuration II. We may characterize this uncertainty as a response function of the photodetector, $D(\tilde{t}_j - t_j)$, where $t_j$ is the photon annihilation time and $\tilde{t}_j$ is the electronic registration time. Thus, the joint photodetection measurement of $D_1$ and $D_2$ can be treated as a convolution between the response functions and the second-order correlation function $G^{(2)}(t_{1} - t_{2})$,
\begin{align}\label{Response-1}
G^{(2)}(\tilde{t}_{1} -\tilde{t}_{2}) = \frac{1}{t^{2}_{c}} \, \int dt_{1} \int  dt_{2} \, G^{(2)}(t_{1} - t_{2})\, D(\tilde{t}_{1} - t_{1}) \, D(\tilde{t}_{2} - t_{2})
\end{align}
where we have normalized the function with the response time, or characteristic time, of the photodetector, $t_{c}$. When using fast detectors, the width of the response functions are much narrower then the temporal width of $G^{(2)}(t_{1} -t_{2})$ so the response functions can be treated as delta functions, $D(\tilde{t}-t) \sim t_{c}\delta(\tilde{t}-t)$. In this extreme case, the measured second-order correlation function will reveal the theoretical expectation of $G^{(2)}(t_{1} - t_{2})$. However, when the response times are larger, the situation is different; especially when the temporal widths of the response functions are much wider then that of $G^{(2)}(t_{1} - t_{2})$. In this case, the second-order correlation function itself can then be treated as a delta function and Eq.~(\ref{Response-1}) turns into the following convolution between the response functions of the two photon counting detectors,
\begin{align}\label{Response-2}
G^{(2)}(\tilde{t}_{1} -\tilde{t}_{2}) = G^{(2)}(0) \, \Big{(}\frac{\tau_{0}}{t^{2}_{c}}\Big{)} \int d\tau \, 
D(\tilde{t}_{1} - \tilde{t}_{2}-\tau) \, D(\tau),
\end{align}
where $\tau = t_{1} - t_{2}$ and we have normalized the delta function with $G^{(2)}(0) \tau_{0}$. The factor $\tau_{0}$ is the second-order correlation time and is a constant defined as the inverse of the bandwidth of the source, $\tau_{0} = 1/\Delta \nu$. Eq.~(\ref{Response-2}) indicates two things: (1) the width of the observed $G^{(2)}(\tilde{t}_{1} -\tilde{t}_{2})$ is now determined by the response function of the photodetectors, which could be significantly broadened compared to the original correlation and (2) the relatively slow response time of the photodetectors may reduce the magnitude of the measured second-order correlation. For Gaussian response functions the convolution yields a Gaussian function of $G^{(2)}(\tilde{t}_{1} -\tilde{t}_{2})$ with a reduced central value of $G^{(2)}(0)$. The reduction factor is roughly $\tau_0 / t_c$.

The above result for slow detectors may not be a problem in the measurement of entangled states; however, it may affect the measurement of a $G^{(2)}$ function for thermal or pseudo-thermal light significantly. The $G^{(2)}$ function of thermal or pseudo-thermal field has two terms, the product of two measured mean photon numbers (trivial part) and the photon number fluctuation correlation (nontrivial part) \cite{ShihBook},
\begin{align}\label{Time-average-22}
G^{(2)}(\tilde{t}_{1} -\tilde{t}_{2}) & = \frac{1}{t^{2}_{c}} 
\int dt_{1} \int  dt_{2} \, \langle n(t_1) n(t_2) \rangle D(\tilde{t}_{1} - t_{1}) \, D(\tilde{t}_{2} - t_{2})\cr
& = \frac{1}{t^{2}_{c}} \int dt_{1} \int  dt_{2} \, \big{[} \langle n(t_1) \rangle \langle n(t_2) \rangle + \langle \Delta n(t_1) \Delta n(t_2) \rangle \big{]} D(\tilde{t}_{1} - t_{1}) \, D(\tilde{t}_{2} - t_{2}). 
\end{align}
The time average over $t_c$ has no effect on the first term (product of mean photon numbers) for a CW thermal or pseudo-thermal field with broad spectrum, because $\langle n(t_1) \rangle \langle n(t_2) \rangle = \bar{n}_1 \bar{n}_2$ is a constant. However, it reduces the magnitude of the second term (photon number fluctuation correlation) significantly when $\tau_{0} \ll t_{c}$,
\begin{align}\label{Time-average-33}
& \frac{1}{t^{2}_{c}} \int dt_{1} \int  dt_{2} \, \langle  \Delta n(t_1) \Delta n(t_2) \rangle D(\tilde{t}_{1} - t_{1}) D(\tilde{t}_{2} - t_{2}) \cr
& \ \ \ \ \ \ \ \ \ \ \ \ \approx \frac{\bar{n}_1 \bar{n}_2}{t^{2}_{c}} \int dt_{1} \int  dt_{2} \, \tau_0 \delta (t_1 - t_2)
D(\tilde{t}_{1} - t_{1}) D(\tilde{t}_{2} - t_{2}) \cr
& \ \ \ \ \ \ \ \ \ \ \ \ \approx   \frac{\bar{n}_1 \bar{n}_2}{t^{2}_{c}} \tau_0 \int d\tau \, D(\tilde{t}_{1} - \tilde{t}_{2}-\tau) \, D(\tau).
\end{align}
For Gaussian response functions, the magnitude of the photon number fluctuation term is thus roughly $\tau_0 / t_c$ times that of the product of mean photon numbers, which may reach $10^{-6}$ (one part of a million) for an X-ray source with $10^{15}$Hz bandwidth ($\tau_0 \sim 10^{-15}$) and a nanosecond photodetector ($t_c \sim 10^{-9}$).

How does one distinguish the relatively small, photon number fluctuations correlation part of the second-order correlation of a thermal field with short coherence time $\tau_0$? The obvious approach is to use fast photodetectors with $t_c \sim \tau_0$. However, even if faster, individual photodetectors are used instead of a full detector array, the state-of-the-art technology has not been able to produce photodetectors fast enough to achieve this for broadband light sources yet. One realistic approach that could utilize current photodetector technology is using a pulsed thermal or pseudo-thermal radiation source with pulse width of $\tau_p$ and recording a single pulse per frame. In this case, the time integral in Eq.~(\ref{Time-average-22}) shall have similar effects on the first term of $G^{(2)}$ (product of mean photon numbers) as it had in Eq.~(\ref{Time-average-33}),
\begin{align}\label{Time-average-44}
& \frac{1}{t^{2}_{c}} \int dt_{1} \int  dt_{2} \langle n(t_1) \rangle \langle n(t_2) \rangle  D(\tilde{t}_{1} - t_{1}) D(\tilde{t}_{2} - t_{2}) \cr
& \ \ \ \ \ \ \ \ \ \ \ \ \approx \frac{\bar{n }_{1} \bar{n}_2}{t^{2}_{c}} \int dt_{1} \int  dt_{2} \, \tau_p \delta (t_1 - t_2) D(\tilde{t}_{1} - t_{1}) D(\tilde{t}_{2} - t_{2}) \cr
& \ \ \ \ \ \ \ \ \ \ \ \ \approx   \frac{\bar{n }_{1} \bar{n}_2}{t^{2}_{c}} \tau_p \int d\tau \, D(\tilde{t}_{1} - \tilde{t}_{2}-\tau) \, D(\tau).
\end{align}
where we have approximated the short pulsed $\langle n(t_1) \rangle \langle n(t_2) \rangle$ as a delta-function such that $\langle n(t_1) \rangle\langle n(t_2) \rangle \sim \bar{n}_1 \bar{n}_2 \tau_p \delta(t_1 - t_2)$. For Gaussian response functions, This results in the pulse being broadened and reduced by a factor of $\tau_p / t_c$. In the case where the pulse width is approximately the same as the coherence time, $\tau_p \approx \tau_0$, all of the photons in the pulse are considered second-order coherent. This results in,
\begin{align}\label{Time-average-55}
\frac{1}{t^{2}_{c}} \int dt_{1} \int  dt_{2} \langle n(t_1) \rangle 
\langle n(t_2) \rangle  D(\tilde{t}_{1} - t_{1}) D(\tilde{t}_{2} - t_{2}) \approx   \frac{\bar{n}_1 \bar{n}_2}{t^{2}_{c}} \tau_0 \int d\tau \, D(\tilde{t}_{1} - \tilde{t}_{2}-\tau) \, D(\tau),
\end{align}
which matches Eq.~(\ref{Time-average-33}), meaning the two terms in the second-order correlation are reduced by the same factor. This allows the photon number fluctuation correlation (intensity fluctuation correlation) to become more distinguishable and is suitable for measuring the second-order correlation of X rays. High-intensity synchrotron X-ray sources are common to use for X-ray imaging. Unfortunately, even though they are pulsed sources, the temporal width of a synchrotron X-ray pulse is still significantly greater than its coherence time, $\tau_p \gg \tau_0$. Previous two-photon X-ray correlation measurements introduced high-resolution monochromators into the X-ray beam of their synchrotron to make the coherence time, $\tau_0$, more comparable to the pulse width $\tau_p$ \cite{XrayHBT1997, XrayHBT1999, XrayHBT2001}. This produced clear correlation that could be used to produce an X-ray ghost image; however, it introduces experimental challenges because of the intensity lost in the monochromators. While the avalanche photodiodes used in the cited demonstrations were able to detect these low light levels, this will prove more difficult with current sensor array (CCD or CMOS) technology. In addition, synchrotron X-ray sources are not necessarily the ideal candidate for the X-ray ghost microscope setups. For most X-ray imaging techniques (along with most other X-ray experiments) it is desirable to have the smallest possible source diameter. This has led to synchrotrons with very small source diameters which is the opposite of the sources with larger angular diameters required for high-resolution X-ray ghost imaging. The ideal setup to fully realize the X-ray ghost microscope would be to pair a pulsed X-ray source with an angular diameter approaching 0.1 rad and a high-resolution monochromator such that the pulse width is comparable to the coherence time, $\tau_p \approx \tau_0$. An X-ray source with a large diameter and short pulsed emission is currently in development at Los Alamos National Laboratory and other institutions.

\section{Summary}

In summary, we have analyzed the working mechanism of the X-ray ghost microscope. By applying Einstein's granularity picture of radiation, which includes X rays, we found that two-photon interference is able to produce an image-forming correlation that forms a lensless X-ray ghost image with high spatial resolution, but no magnification. Utilizing a high-energy ($>$ 20 keV) source with a relatively large angular diameter, a ghost image with sub-nanometer resolution could be obtained. Furthermore, either with the addition of focusing X-ray optics or the help of a scintillator and an optical lens system, the primary ghost image can be mapped onto a secondary image plane with significant magnification but reduced (but still high) resolution. Because the primary ghost image is along a different beam than the physical object, the focusing X-ray optics or scintillator can be placed anywhere along the beam, allowing for high-resolution, magnified imaging of the deep internal structure of an object without direct physical constraints. Linearly scanning the secondary imaging setup along the propagation axis, a 3-D image can be obtained with Linear Ghost Tomography. We have also found that a short-pulsed X-ray source and measuring a single pulse per frame is preferred for observing the two-photon interference induced intensity fluctuation correlation of X rays due to the wide bandwidth of the source. Even with measuring a single pulse per frame, in order to achieve a higher degree of second-order correlation it is preferred to have a more monochromatic beam such that the coherence time, $\tau_0$, is more comparable to the pulse width, $\tau_p$. In addition to this, the desire for a large angular diameter makes synchrotron X-ray sources less favorable as the resolution of the primary ghost image would be greatly reduced. As sensor arrays and X-ray sources see technological advancements, this X-ray ghost microscope will open up new capabilities that would be of interest to the fields of physics, material science, and medical imaging.

\section*{Funding}

Los Alamos National Laboratory.

\section*{Acknowledgments}

The authors thank J. N. Sprigg and B. Joshi for helpful discussions.

\section*{Disclosures}

The authors declare no conflicts of interest related to this article.


\begin{thebibliography}{999}
\bibitem{footnoteXFEL} Additionally, X-ray free-electron lasers (XFELs) do not produce X rays in a thermal state, so type (I) ghost imaging is not possible, unless the coherent X-ray beam is thermalized into a pseudo-thermal state; although, type (II) ghost imaging is possible when a beam diffuser is used to introduce speckles \cite{KimXR}.
\bibitem{footnoteBS} This is trivial when using visible light, but becomes more difficult with high-energy X-rays. So far, Laue diffraction of crystals have been used with success, but advancement in kinoform X-ray beam splitters may prove useful in the future \cite{XRBS}.
\bibitem{footnote3D} A benefit of ``linear'' ghost tomography as opposed to ``rotational'' ghost tomography is the theoretical capability of having a 3-D array of photodetectors such that each 2-D plane captures a slice of the object simultaneously, removing the need to move the object or photodetectors. Of course the technology of 3-D photodetector arrays does not exist yet, but initial assessments have recently been completed for multilayer ultrafast silicon sensor development \cite{3Darray}.
\end{thebibliography}
\end{document}